\documentclass[sigconf, nonacm, screen]{acmart}

\usepackage{pifont}
\usepackage{subfig}
\usepackage{amsthm}
\usepackage{csquotes}
\usepackage{mathtools}
\usepackage{listings}
\usepackage{multirow}
\usepackage{tikz}
\usepackage{balance}

\theoremstyle{definition}

\usepackage{pifont}
\usepackage[font=small,labelfont=bf,tableposition=top]{caption}
\DeclareCaptionLabelFormat{andtable}{#1~#2  \&  \tablename~\thetable}
\usepackage[ruled,vlined,noend]{algorithm2e}
\usepackage{listings}
\usepackage{natbib}
\usepackage{graphicx}

\definecolor{lightgray}{gray}{0.5}
\def\MM#1{\ifmmode#1\else\mbox{$#1$}\fi}
\usepackage[export]{adjustbox}
\usepackage{enumitem}
\usepackage{makecell}
\usepackage{soul}

\def\diogo#1{\hphantom{{\sc \textcolor{brown}{Diogo: }}{\sf \textcolor{red}{#1}}}}

\def\tododiogo#1{\hphantom{{\sc \textcolor{brown}{TODO-Diogo: }}{\sf \textcolor{red}{#1}}}}

\DeclareMathDelimiter{(}{\mathopen} {operators}{"28}{largesymbols}{"00}
\DeclareMathDelimiter{)}{\mathclose}{operators}{"29}{largesymbols}{"01}
    
\usepackage{lineno}

\begin{document}

\title{Selectivity Estimation of Inequality Joins In Databases}

\author{Diogo Repas}
\orcid{0000-0003-4983-0032}
\affiliation{%
  \institution{Universit\'e libre de Bruxelles (ULB)}
  \city{Brussels}
  \country{Belgium}
}
\email{diogo.seca.repas.goncalves@ulb.be}

\author{Zhicheng Luo}
\orcid{0000-0001-5109-3700}
\affiliation{%
  \institution{Universit\'e libre de Bruxelles (ULB)}
  \city{Brussels}
  \country{Belgium}
}
\email{zhicheng.luo@ulb.be}

\author{Maxime Schoemans}
\affiliation{%
  \institution{Universit\'e libre de Bruxelles (ULB)}
  \city{Brussels}
  \country{Belgium}
}
\email{maxime.schoemans@ulb.be}

\author{Mahmoud Sakr}
\orcid{0000-0002-1825-0097}
\affiliation{%
  \institution{Universit\'e libre de Bruxelles (ULB)}
  \city{Brussels}
  \country{Belgium}\\
  \institution{Ain Shams University}
  \city{Cairo}
  \country{Egypt}
}
\email{mahmoud.sakr@ulb.be}

\begin{abstract}
Selectivity estimation refers to the ability of the SQL query optimizer to estimate the size of the results of a predicate in the query. It is the main calculation, based on which the optimizer can select the cheapest plan to execute. While the problem is known since the mid 70s, we were surprised that there are no solutions in the literature for the selectivity estimation of inequality joins. By testing four common database systems: Oracle, SQL-Server, PostgreSQL, and MySQL, we found that the open-source systems PostgreSQL and MySQL lack this estimation. Oracle and SQL-Server make fairly accurate estimations, yet their algorithms are secret. This paper thus proposes an algorithm for inequality join selectivity estimation. The proposed algorithm has been implemented in PostgreSQL and sent as a patch to be included in the next releases.
\end{abstract}

\maketitle

\section{Introduction}

Query optimization is the overall process of generating the most efficient query plan given an SQL statement. The query optimizer, responsible for this process, applies equivalence rules to reorganize and merge the operations in the query to find the fastest execution plan and feeding it to the executor. It examines multiple access methods, such as sequential table scans or index scans, different join methods such as nested loops and hash joins, different join orders, sub-query normalization, materialized views, and other possible transformations. Starting with a naively generated query plan, the optimizer generates a set of equivalent plans. To choose the most efficient plan, almost all systems adopt a cost-based approach, which roots back in the architecture of System R \cite{systemR} and Volcano/Cascades \cite{Cascades,Volcano}.

In cost-based query optimization, the optimizer estimates the cost of the alternative query plans and chooses the plan with minimum cost. The cost is estimated in terms of the CPU and I/O resources that the query plan will use. A central component in cost estimation is the Selectivity Estimation (SE). SE collects statistics for all attributes in a relation, such as data distribution histograms, most common values, null percentage, etc. These statistics are then used during planning time to estimate the number of tuples generated by a predicate in the query. A smaller selectivity value means a smaller size for intermediate results, which is favorable for a more efficient execution. The cost-based optimizer thus reorders the selection and join predicates to quickly reduce the sizes of intermediate results.

Since, in general, the cost of each operator depends on the size of its input relations, it is important to provide good estimations of their selectivity, that is, of their result size, to the query optimizer \cite{Pitoura2009}.
Inaccurate selectivity estimations can lead to inefficient query plans being chosen, sometimes leading to orders of magnitude longer execution times\cite{Lan2021ASO}.

There is a trade-off between the size of the stored statistics and the complexity of the estimation algorithm on the one hand, and the estimation accuracy on the other. Recent research thus focuses on using machine learning methods to capture the data distribution into compact models. While there are good results in this research direction \cite{10.14778/3461535.3461552}, common relational database systems continue to use traditional statistics structures, mostly based on histograms. A histogram can be used as a discrete approximation of the probability density function of an attribute. 

Despite the popularity of histograms, there is a lack of theory on how to use them in estimating inequality join selectivity.
This paper aims at filling this gap, and presents the following main contributions:
\begin{itemize}
    \item A novel algorithm for join selectivity estimation of inequality operators using histogram statistics
    \item The implementation of this algorithm in PostgreSQL both for scalar inequality joins, as well as for multiple operators of range types
    \item An extension of the algorithm that also takes advantage of additional statistics, when available.
    \item The proposed algorithm has been implemented in PostgreSQL and submitted as a patch for inclusion in a future release\footnote{\url{https://github.com/DRepas/postgres/tree/rangejoinsel}}
\end{itemize}

Section \ref{section:review} starts by reviewing existing work in selectivity estimation. A running example to be used throughout the paper is given in Section \ref{section:example}. Then, some definitions, notation and terminology are introduced in Section \ref{section:hist_stats}.
The algorithm presented in this paper consists of mapping the problem of selectivity estimation to a probability theory problem, as described in Section \ref{section:formal}.
The algorithm of join selectivity estimation is developed in Section \ref{section:impl}. The section also develops a way to incorporate null values and Most Common Values (MCV) statistics in the estimation model. A mapping of this model for several range operators is also presented. 
Finally, an implementation in PostgreSQL and the experimental evaluation are provided in Section \ref{section:experiments}.

\section{Review of Related Work}
\label{section:review}

A survey on DBMS query optimizer has been proposed in 2021 \cite{Lan2021ASO}, which categorizes cardinality estimation methods into synopsis-based methods, sampling-based methods, and learning-based methods. 
Many learning-based methods \cite{10.1145/3318464.3389741, DBLP:journals/corr/abs-1809-00677, DBLP:journals/corr/abs-1905-04278} have been proposed in recent years and show better accuracy than traditional methods. 
But there are still many missing parts to be solved to put them into real systems, such as the cost of model training and updating, and the black-box property of learning algorithms \cite{10.14778/3461535.3461552}.
Sampling-based methods estimate selectivity by executing a (sub)query on samples collected from tables, whose accuracy depends on the degree to which the samples fit the original data distribution \cite{Lan2021ASO}.
These methods, however, suffer from a high cost of storage and retrieval time, especially when the tables are very large. 
Another limitation of sampling-based methods is that they currently only support equality join selectivity estimation \cite{Lan2021ASO}.
Histograms, as a form of synopsis-based methods, have been extensively studied \cite{Lan2021ASO,10.1145/3318464.3389741} and are widely adopted in common database systems \cite{Cormode2012} for the purpose of selectivity estimation, including MySQL, PostgreSQL, Oracle, and SQL Server \cite{mysql, postgresql, oracle, sqlserver_statistics}.


MySQL uses two histogram types for selectivity estimation \cite{mysql}. 
One is the singleton histogram which stores the distinct values and their cumulative frequency.
Another is the equi-depth histogram, called equi-height in MySQL documentation. This histogram stores the lower and upper bounds, cumulative frequency, and the number of distinct values for each bucket.
However, the usage of histograms is limited to restriction selectivity estimation \cite{mysql}, i.e., the selection operator. For join selectivity estimation, MySQL naively returns a constant: 0.1 for equality joins and 0.3333 for inequality joins \cite{mysql_default_selectivity}. The following is the excerpt of the MySQL source code in which these constants are defined \cite{mysql_source_2014}:

\begin{lstlisting}[language=C++]
/// Filtering effect for equalities: col1 = col2
#define COND_FILTER_EQUALITY 0.1f
/// Filtering effect for inequalities: col1 > col2
#define COND_FILTER_INEQUALITY 0.3333f
\end{lstlisting}

PostgreSQL also uses histogram as optimizer statistics \cite{postgresql}. 
By analyzing its manual \cite{postgresql_statistics_usage}, as well as its source code, it uses equi-depth histograms. In contrast to MySQL, the number of distinct values in each bucket is not stored. For this reason, PostgreSQL does not use these histogram statistics in estimating equi-join selectivity. It rather uses a singleton histogram of Most Common Values (MCV) \cite{postgresql_statistics_usage}.
As for inequality join selectivity estimation ($<$, $\leq$, $>$, $\geq$), a default constant value of $0.3333~$ is returned \cite{postgre_default_selectivity}.

The following is an excerpt of the PostgreSQL source code in which these constants are defined \cite{postgres_source_2004}:

\begin{lstlisting}[language=C]
/* default selectivity estimate for equalities such as  "A = b"*/
#define DEFAULT_EQ_SEL	0.005
/* default selectivity estimate for inequalities such as "A < b"*/
#define DEFAULT_INEQ_SEL  0.3333333333333333
\end{lstlisting}

Oracle Database uses three types of histograms to capture the data distribution of a relation's attribute \cite{oracle}: singleton histograms (referred to as frequency histogram and top frequency histogram in the official documentation), equi-depth histogram (referred to as height-balanced in the official documentation), and hybrid histogram (a combination of equi-depth and frequency histograms). 
The type of histogram is determined based on specific criteria to fit different situations.
The official documentation \cite{oracle} also states some factors behind their selectivity estimation algorithms, such as endpoint numbers (the unique identifier of a bucket, e.g., the cumulative frequency of buckets in frequency and hybrid histograms) and values (the highest value in a bucket), and whether column values are popular (an endpoint value that appears multiple times in a histogram) or non-popular (every column value that is not popular).
However, the details of these estimation algorithms are not published. Few online articles, in the form of hacker blogs, did experimental analyses to guess how selectivity estimation works in Oracle Database but didn't yield a clear algorithm,   \cite{Lewis2006, oracle_join}.

SQL-Server is another popular closed-source DBMS.
Due to its proprietary nature, implementation details are scarce.
According to the official documentation \cite{sqlserver_statistics}, a proprietary kind of histogram with a density vector associated is built in three steps for each attribute.
The official documentation \cite{sqlserver_CE, sqlserver_ce_white_paper} describes four core assumptions for the selectivity estimation: independence when no correlation information is available, uniformity in histogram bins, inclusion when filtering a column with a constant, and containment when joining distinct values from two histograms \cite{Bruno2002}.
Although the white paper \cite{sqlserver_ce_white_paper} is a publication from SQL Server that deals with the problem of selectivity estimation, it does not explain the algorithm used for join selectivity. Similar to Oracle, the implemented algorithm is a secret.

To identify if any informed selectivity estimation is taking place when performing inequality joins in SQL-Server and Oracle, we have performed the following experiment. Two different attributes, T1 and T2, with 1000 and 200 rows respectively, were randomly generated by sampling the range [0, 100] uniformly. They were then joined using the $<$ (less than) operator. Both databases made a quite accurate selectivity estimation of this inequality join. Oracle Database had an estimation error of 3\%, and SQL Server had a smaller error of just 0.29\%. As such, we know that both systems implement good estimation algorithms.

In conclusion, although learning-based methods have become a popular research direction for selectivity estimation in recent years, histograms are still the most commonly used statistics in existing DBMS for this purpose. The recurring types of used histogram statistics are equi-depth histograms approximating the distribution of values, and singleton histograms of Most Common Values. As our investigation indicates, MySQL and PostgreSQL don't have algorithms implemented for join selectivity estimation, and they use predefined constants.
On the other hand, popular commercial DBMS (SQL-Server and Oracle) have implemented some algorithms based on the histograms, but we couldn't find any source describing them. This paper addresses this gap by proposing such an algorithm.

\section{Preliminaries}
\label{section:preliminaries}

This paper presents a formal model to reason about two different selectivity estimation types:
\begin{itemize}
    \item Restriction selectivity estimation: when one of the sides of the operator is an attribute of a relation and the other is a constant value.
    \begin{itemize}
        \item Example: SELECT * FROM R1 WHERE R1.X < 100
    \end{itemize}
    \item Join selectivity estimation: when both sides of the operator are attributes of different relations.
    \begin{itemize}
        \item Example: SELECT * FROM R1, R2 WHERE R1.X < R2.Y
    \end{itemize}
\end{itemize}

Selectivity estimation of operations where both sides of the operator are attributes of the same relation, with no join or Cartesian product involved (Example: SELECT * FROM R1 WHERE R1.x < R1.y), is not addressed by this paper. 

The selectivity of an operator is the fraction of values kept in the result after selection.
In the case of restriction selectivity, the denominator is the input relation's size.
In the case of join selectivity, the denominator is the input relations' Cartesian product size (their sizes multiplied). This fraction can be interpreted as the probability that a given randomly selected tuple from the input relation, or from the Cartesian product of input relations in the case of joins, is selected by the operator being considered.

The focus of the next sections will be on the restriction selectivity estimation of the \emph{less than (<)} operator. The restriction selectivity estimation of all scalar inequality operators will be derived from this initial estimation. We will also build/generalize on it to develop the join selectivity estimation. This restriction selectivity estimation in the next section is already implemented by all common database systems, thus not a novel contribution of this work. We however formulate it as a probability problem, and develop the join selectivity estimation on top of it, to maximize the code reuse in these systems.

The attributes being restricted or joined will be treated as random variables that follow a distribution modeled by a Probability Density Function (PDF) and/or a Cumulative Distribution Function (CDF). 

\subsection{Running Example}
\label{section:example}

For demonstration purposes, relations R1 and R2 will be used throughout this paper. 
For each relation, 12 integers were manually selected to cover as many corner cases as possible when using equi-depth histograms (introduced in section \ref{section:hist_stats}), such as skew and common bin boundaries.

$R1.X= \{10, 11, 12, 20, 21, 22, 24, 25, 30, 35, 38, 45\}$

$R2.Y= \{15, 16, 17, 20, 30, 35, 38, 39, 40, 42, 45, 50\}$

\subsection{Histogram Statistics}
\label{section:hist_stats}

Histograms are commonly used to approximate the PDF of an attribute by grouping the values in uniform bins. Each bin is an interval of the form \(B_j = ]hist_j, hist_{j+1}]\), where $hist_i$ are values from the domain of the attribute. It is important to note that the side on which the interval is open or closed is not relevant for the purposes of this paper, as all estimations correspond to integration over a continuous domain, where singular points do not affect the final result. By defining a bin this way (using intervals), this paper is restricting itself to domains where total order exists. This excludes categorical data, for which the idea of an equi-depth histogram does not apply.

Let \(H_j(X) = P(X \in B_j)\) be the fraction of values of the attribute \(X\) that is represented by the histogram bin \(B_j\), i.e., the height/depth of the histogram bin:
\begin{itemize}
    \item \emph{Singleton histograms} are such that each bin refers to the frequency of a single element. Typically used to collect Most Common Values statistics (introduced in section \ref{section:other_stats}).
    \item \emph{Equi-width histograms} are such that each bin has the same width. That is, \(hist_{j+1} - hist_j\) is constant.
    \item \emph{Equi-depth histograms} are such that each bin has the same depth (height) but varying width. That is \(\forall j, P(X \in B_j) = H_j(X) = \frac{1}{n}\), where \(n\) is the total number of bins.
\end{itemize}

For selectivity estimation, equi-depth histograms are favoured because the resolution of the histogram adapts to skewed value
distributions. Typically, the histogram is smaller than the original data. Thus, it cannot represent the true distribution entirely and some assumptions are induced, e.g., uniformity on a single attribute, and independence assumption among different attributes. The use of equi-depth histograms is also motivated by their prevalence in current RDBMS. These are usually constructed through the use of random sampling of the original relations.

\begin{figure}[ht]
\centering
    \begin{tikzpicture}[scale=0.9]
    \draw (2, 1) node[left] {$R1.X$} -- (9, 1);
    \draw (2, 1cm - 4pt) node[below] {$10$} -- (2, 1cm + 4pt);
    \draw (4, 1cm - 4pt) node[below] {$20$} -- (4, 1cm + 4pt);
    \draw (5, 1cm - 4pt) node[below] {$25$} -- (5, 1cm + 4pt);
    \draw (9, 1cm - 4pt) node[below] {$45$} -- (9, 1cm + 4pt);
    \draw (3, 2) node[left] {$R2.Y$} -- (10, 2);
    \draw (3, 2cm - 4pt) node[below] {$15$} -- (3, 2cm + 4pt);
    \draw (4, 2cm - 4pt) node[below] {$20$} -- (4, 2cm + 4pt);
    \draw (7.8, 2cm - 4pt) node[below] {$39$} -- (7.8, 2cm + 4pt);
    \draw (10, 2cm - 4pt) node[below] {$50$} -- (10, 2cm + 4pt);
    \end{tikzpicture}
\caption{Equi-depth histograms of R1.X and R2.Y with 3 bins each.}
\label{fig:equidepth_hist}
\end{figure}
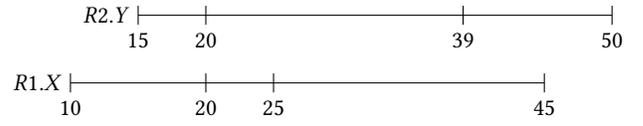

For demonstration purposes, Figure \ref{fig:equidepth_hist} shows a graphical representation of equi-depth histograms for the attributes $R1.X$ and $R2.Y$ of the running example. For both histograms, there are three bins, meaning that the fraction accounted for by each bin is \(\frac{1}{3}\).
For attribute $R1.X$, \(hist_X = [10, 20, 25, 45]\), which means that \({B_X}_0 = [10, 20]\), \({B_X}_1 = ]20, 25]\), and \({B_X}_2 = ]25, 45]\).
For attribute y, \(hist_Y = [15, 20, 38, 50]\), which means that \({B_Y}_0 = [15, 20]\), \({B_Y}_1 = ]20, 39]\), and \({B_Y}_2 = ]39, 50]\).

Using histograms as a statistical representation of attributes involves the following implicit assumptions:
\begin{itemize}
    \item The data is distributed uniformly inside each bin
    \item The histograms are complete (they account for all the data points), that is:
    \begin{itemize}
        \item \(hist_0 = min(X)\)
        \item \(hist_n = max(X)\)
    \end{itemize}
\end{itemize}
In practice, these two assumptions do not strictly hold. The data is usually not uniformly distributed inside each bin. The more bins used in the histograms, the smaller the error introduced by this assumption. Database systems, e.g., PostgreSQL, typically create the histogram using a random sample of the attribute values, especially when the number of tuples is too large. The assumption of completeness of the histogram might be broken in the presence of sampling.
When the sample is representative of the underlying data, the estimation is still fairly accurate.

Given the equi-depth histogram of an attribute \(X\), with \(n\) bins, one can derive its approximate PDF and CDF as shown next. Let $f_X(c)$ and $F_X(c)$ denote the PDF and the CDF of X, respectively, at a given value $c$ then:  
\begin{equation}
\label{eq:discrete_pdf}
f_X(c) = \begin{cases} 
    0 & c < hist_0 \\
    \frac{1}{n} \frac{1}{(hist_{j+1} - hist_j)} & c \in B_j, 0 \leq j < n \\
    0 & c \geq hist_n
\end{cases}
\end{equation}

\begin{equation}
\label{eq:discrete_cdf}
F_X(c) = \begin{cases} 
    0 & c < hist_0 \\
    \frac{1}{n} (j + \frac{c - hist_j}{hist_{j+1} - hist_j}) & c \in B_j, 0 \leq j < n \\
    1 & c \geq hist_n
\end{cases}
\end{equation}
When \(c \in B_j, 0 \leq j < n\), formula \ref{eq:discrete_pdf} is derived from the definition of equi-depth histogram, where each bin represents \(H_j(X) = \frac{1}{n}\) of the data, spread over a width of \(hist_{j+1} - hist_j\). 

Formula \ref{eq:discrete_cdf} is derived from the following:

\[ F_X(c) = \int_{-\infty}^{c} f_X(x) \,dx = \]
\[ \int_{-\infty}^{hist_0} f_X(x) \,dx + \sum_{i = 0}^{j-1} \int_{hist_i}^{hist_{i+1}} f_X(x) \,dx + \int_{hist_j}^{c} f_X(x) \]
where,

\[ \int_{-\infty}^{hist_0} f_X(x) \,dx = 0 \]
\[ \sum_{i = 0}^{j-1} \int_{hist_i}^{hist_{i+1}} f_X(x) \,dx = \sum_{i = 0}^{j-1} H_j(X) = \sum_{i = 0}^{j-1} \frac{1}{n} = \frac{j}{n} \]
\[ \int_{hist_j}^{c} f_X(x) = H_j(X) \frac{c - hist_j}{hist_{j+1} - hist_j} \,dx = \frac{1}{n} \frac{c - hist_j}{hist_{j+1} - hist_j} \]
This last formula performs linear interpolation within the bin where \(c\) is contained, thus assuming a uniform distribution of values within the bin. The assumption that the histogram is complete is reflected in substituting the infinite bounds by $hist_0, hist_n$. 
In the running example, the PDF and CDF of $R1.X$ can be derived from the formulas presented in this section as follows:

\begin{equation*}
f_X(c) = \begin{cases} 
    0 & c < 10 \\
    \frac{1}{30} & 10 \leq c < 20 \\
    \frac{1}{15} & 20 \leq c < 25 \\
    \frac{1}{60} & 25 \leq c < 45 \\
    0 & 45 \leq c
\end{cases}
\end{equation*}

\begin{equation*}
F_X(c) = \begin{cases} 
    0 & c < 10 \\
    \frac{1}{30} c - \frac{1}{3} & 10 \leq c < 20 \\
    \frac{1}{15} c - 1 & 20 \leq c < 25 \\
    \frac{1}{60} c + \frac{1}{4} & 25 \leq c < 45 \\
    1 & 45 \leq c
\end{cases}
\end{equation*}

\section{A Formal Model for Selectivity Estimation}
\label{section:formal}

We first start by formalizing the problem of restriction selectivity estimation for the \emph{Less Than (<)} operator. Suppose the goal of estimating the selectivity of the following operation (expressed in SQL):

\begin{lstlisting}[language=SQL]
SELECT *
FROM R1
WHERE R1.X < c
\end{lstlisting}
where c is a constant. Treating the attribute $R1.X$ as a random variable \(X\), estimating the selectivity of the above operation is equivalent to finding \(P(X < c)\). Given the PDF or the CDF of \(X\), \(f_X\) or \(F_X\), respectively, the selectivity of the operation above can be formalized as:
\begin{equation}
\label{eq:cdf}
    P(X < c)= \int_{-\infty}^{c} f_X(x) \,dx= F_X(c)
\end{equation}

Suppose now that the goal is estimation the join selectivity for the \emph{Less Than (<)} operator. That is, we want to  estimate the selectivity of the following operation (expressed in SQL):
\begin{lstlisting}[language=SQL]
SELECT *
FROM R1, R2
WHERE R1.X < R2.Y
\end{lstlisting}
Treating the attributes $R1.X$ and $R2.Y$ as random variables \(X\) and \(Y\), respectively, estimating the selectivity of the above operation can be formulated as finding \(P(X < Y)\).

Consider the joint distribution of \(X\) and \(Y\), \(P(X,Y)\).
The probability that a sample (a, b) taken at random from the Cartesian product of the values in $R1.X$ and $R2.Y$ can be defined as follows:
\[
    \forall a \in R1.X, b \in R2.Y, P(X= a, Y= b)= P(X=a) \times P(Y=b)
\]
or equivalently:
\[
    P(X, Y) = P(X) P(Y)
\]
which is the definition of independent random variables. Note that when a Cartesian product is involved, either explicitly as in the SQL statement above, or implicitly through a join clause, the two variables are independent. 

Given a joint PDF of \(X\) and \(Y\), \(f_{X,Y}\).
With \(X\) and \(Y\) being independent, it is known that \(f_{X,Y}(x,y) = f_X(x)f_Y(y)\), with \(f_X\) and \(f_Y\) the PDFs of \(X\) and \(Y\), respectively.
Considering \(F_X\) to be the CDF of \(X\), the selectivity of the \emph{less than (<)} operator can be formalized as follows:

\begin{multline}
\label{eq:join}
P(X < Y) = \\
\int_{-\infty}^{+\infty} \int_{-\infty}^{y} f_{X,Y}(x,y) \,dx \,dy = \\
\int_{-\infty}^{+\infty} \int_{-\infty}^{y} f_X(x) f_Y(y) \,dx \,dy = \\
\int_{-\infty}^{+\infty} (\int_{-\infty}^{y} f_X(x) \,dx) f_Y(y) \,dy = \\
\int_{-\infty}^{+\infty} F_X(y) f_Y(y) \,dy \\
\end{multline}
This formula thus presents a solution for estimating the join selectivity estimation. Next, we discuss how to translate it into an algorithm.

\section{Implementation in a Database System}
\label{section:impl}
In RDBMS implementations, histograms are used as a discrete approximation of the PDF and CDF of attributes. This section maps the theory above into an implementable solution in databases using equi-depth histograms.
\subsection{Selectivity Estimation}

\subsubsection*{Restriction selectivity estimation}
\label{discrete:restriction_estimation}

Recall that restriction selectivity estimation is about estimating the selectivity of a predicate in the following form:

\begin{lstlisting}[language=SQL]
SELECT *
FROM R1
WHERE R1.X < c
\end{lstlisting}

As described in section \ref{section:formal}, restriction selectivity estimation can be calculated using the CDF of \(X\) (equation \ref{eq:cdf}).
Deriving the CDF from an equi-depth histograms is presented in equation \ref{eq:discrete_cdf}.
To find the bin, \(B_j\), where \(c\) is contained, one can perform a binary search over the histogram boundaries.

Algorithm \ref{alg:restriction_sel_estimation} illustrates the estimation of restriction selectivity. The \(hist\) array represents the equi-depth histogram, stored as an ordered array of bin boundaries. The function binary\_search returns the greatest index in this array that is less than or equal to a given constant, effectively finding the bin where such constant falls into. The rest of the algorithm computes the CDF at c using equation \ref{eq:discrete_cdf}.

\begin{algorithm}
    \caption{Restriction selectivity estimation for the expression R1.X < c}
    \label{alg:restriction_sel_estimation}
    \KwIn{\(hist\) an array of length \(n\) representing the equi-depth histogram of R1.X, \(c\) the scalar literal in the query 
    }
    \KwOut{the estimated selectivity}
    
    \tcc{Identify preceding whole bins}
    j $\leftarrow$ binary\_search(hist, c) \;
    
    \tcc{Corner cases}
    \If{j < 0}{\Return{0}}
    \If{j >= n - 1}{\Return{1}} 
    
    \tcc{Estimate using preceding bins}
    selectivity $\leftarrow$ j / (n - 1) \;
    
    \tcc{Adjust using linear interpolation}
    selectivity += (c - hist[j]) / (hist[j+1] - hist[j]) / (n - 1) \;
    
    \Return{selectivity}
\end{algorithm}

Using the running example, and the following query:
\begin{lstlisting}[language=SQL]
SELECT *
FROM R1
WHERE R1.X < 30
\end{lstlisting}
this query yields 8 rows, which corresponds to a selectivity of \(\frac{2}{3}\).
Using Algorithm \ref{alg:restriction_sel_estimation} with the histograms presented in the running example, the number 30 will be found in \({B_X}_2\), meaning that the estimated selectivity will be \(\frac{2}{3} + \frac{1}{3} \frac{30-25}{20} = \frac{9}{12}\).
After multiplying by the attribute's cardinality (12), we get the estimated row count of 9, which is a close estimate to the actual result size.

\begin{figure}[ht]
\centering
\includegraphics[width=0.7\linewidth]{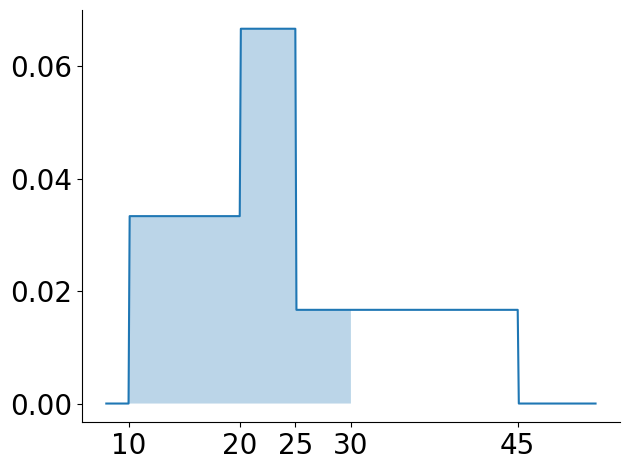}
\caption{Restriction Selectivity Estimation of \(R1.X < 30\)}
\label{fig:integralPDF}
\end{figure}

Figure \ref{fig:integralPDF} shows a graphical depiction of the PDF of the $R1.X$, which is directly obtainable from its equi-depth histogram.
The integral in equation \ref{eq:discrete_cdf}, as well as the estimation calculated above using Algorithm \ref{alg:restriction_sel_estimation}, correspond to the highlighted area in the figure.

\subsubsection*{Join selectivity estimation}
Join selectivity estimation is mapped into a double integral involving the two PDFs. Equation \ref{eq:join} illustrates that join selectivity can be estimated by using the CDF of \(X\) and the PDF of \(Y\). These can be calculated using equations \ref{eq:discrete_pdf} and \ref{eq:discrete_cdf}.

\begin{figure}[ht]
\centering
\includegraphics[width=0.7\linewidth]{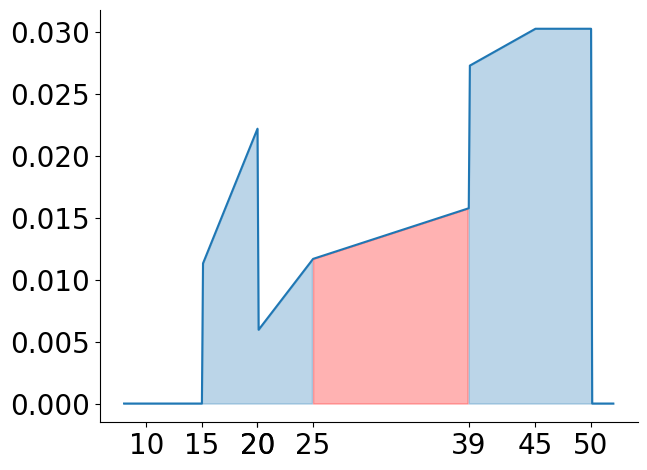}
\caption{Piece of product \(F_X \cdot f_Y\) used for Join Selectivity Estimation}
\label{fig:integral_trapezoid}
\end{figure}

The CDF of $X$ is linear piece-wise, each piece is defined in a bin of $X$'s histogram. The PDF of $Y$ is a step function, i.e., constant piece-wise, where each piece is defined in a bin of $Y$'s histogram. This leads to the conclusion that their product, which is needed in equation \ref{eq:join}, is a linear piece-wise function, with every piece being defined in an intersection of $X$ and $Y$'s bins (see Figure \ref{fig:integral_trapezoid} for a graphical depiction of this using the running example introduced in section \ref{section:example}). By merging the bounds of the two histograms of $X$ and $Y$ in a single sorted array, \(sync\), the intersections of $X$ and $Y$'s bins will be of the form \([sync_j, sync_{j+1}[\).

From equation \ref{eq:discrete_pdf}, we know that \(f_Y\) is 0 for all values until first value of \(hist_Y\), and it is 0 after the last value of \(hist_Y\).
From equation \ref{eq:discrete_cdf}, we know that \(F_X\) is 0 for all values until first value of \(hist_X\), and it is 1 after the last value of \(hist_X\).

Analyzing the product \(F_X \cdot f_Y\) piece-wise:

\begin{equation}
\label{eq:discrete_join_pieces}
(F_X \cdot f_Y)(c) = \begin{cases} 
    0 & c \leq max({hist_X}_0, {hist_Y}_0) \\
    F_X(c) \cdot f_Y(c) & max({hist_X}_0, {hist_Y}_0) < c \\
    & < min({hist_X}_{n_X-1},{hist_Y}_{n_Y-1}) \\
    f_Y(c) & {hist_X}_{n_X-1} \leq c < {hist_Y}_{n_Y-1} \\
    0 & c \geq max({hist_X}_{n_X-1},{hist_Y}_{n_Y-1})
\end{cases}
\end{equation}

Given that the product, \(F_X \cdot f_Y\), in equation \ref{eq:join}, is linear in each interval of the form \([sync_j, sync_{j+1}[\), equation \ref{eq:join} can be discretized as follows:

\begin{multline}
\label{eq:discrete_join}
P(X < Y) = \\
\int_{-\infty}^{+\infty} (F_X \cdot f_Y)(y) \,dy = \\
\sum_{k=0}^{n_X + n_Y - 1} \int_{sync_k}^{sync_{k+1}} (F_X \cdot f_Y)(y) \,dy = \\
\sum_{k=0}^{n_X + n_Y - 1} \frac{(F_X \cdot f_Y)(sync_k) + (F_X \cdot f_Y)(sync_{k+1})}{2} (sync_{k+1} - sync_k) \\
\end{multline}

To maximize code re-use, it is possible to reorganize this equation into a sum of the CDFs of both $X$ and $Y$, i.e., so that we can reuse Algorithm \ref{alg:restriction_sel_estimation}. The following is derived directly from equation \ref{eq:cdf}: 
\begin{multline}
f_Y(sync_k) (sync_{k+1} - sync_k)= \\
f_Y(sync_{k+1}) (sync_{k+1} - sync_k) = \\
\int_{sync_k}^{sync_{k+1}} f_Y(y) \,dy = \\
F_Y(sync_{k+1}) - F_Y(sync_k) \\
\end{multline}
the first two steps rely on the fact that \(f_Y\) is constant in the interval \([sync_k, sync_{k+1}[\).

Equation \ref{eq:discrete_join} can now be re-written using only CDFs of $X$ and $Y$ as follows:

\begin{multline}
\label{eq:discrete_join_inter}
P(X < Y)= \\
\sum_{k=0}^{n_X + n_Y - 1} \frac{(F_X \cdot f_Y)(sync_k) + (F_X \cdot f_Y)(sync_{k+1})}{2} (sync_{k+1} - sync_k) = \\
\frac{1}{2} \sum_{k=0}^{n_X + n_Y - 1} (F_X(sync_k) + F_X(sync_{k+1})) (F_Y(sync_{k+1}) - F_Y(sync_k)) \\
\end{multline}

\begin{figure}[ht]
\centering
\includegraphics[width=0.7\linewidth]{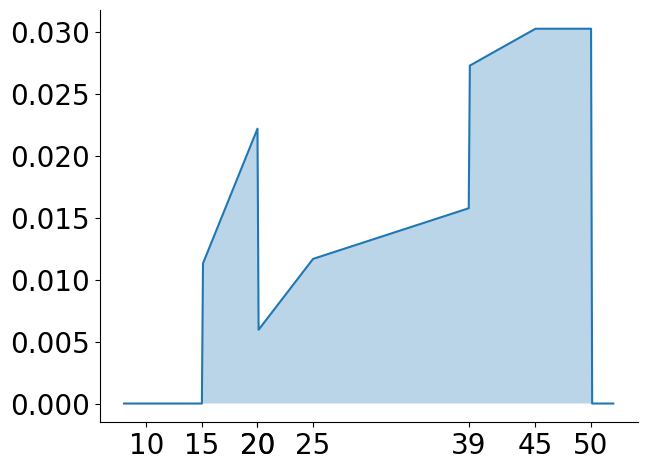}   
\caption{Product \(F_X \cdot f_Y\) used for Join Selectivity Estimation}
\label{fig:integral_product}
\end{figure}

Algorithm \ref{alg:join_sel_estimation} illustrates the estimation of join selectivity of the \emph{less than (<)} operator using this equation. 
It thus has the advantage of re-using algorithm \ref{alg:restriction_sel_estimation} to compute \(F_X\) and \(F_Y\). 
The creation of the \(sync\) array in Algorithm \ref{alg:join_sel_estimation} comes at the expense of time and space of O(\(n_X + n_Y\)), for duplicating and merging the two sorted histograms. 
To optimize this, the two histograms can be scanned in parallel, without the need to materialize the sync array.
This also allows for further optimization. 
The algorithm only needs to iterate over the overlapping region of both histograms. All the partial sums before that will be zero, as can be verified in Figure \ref{fig:integral_product}. After the overlapping region the remaining partial sums are equal to what is left of the histogram of $Y$, \(1 - {cur_F}_Y\), because all remaining values of Y will be greater than the maximum value in $X$.
We adopt these optimizations in our implementation, yet we omit them here for the clarity of presentation.
\begin{algorithm}
    \caption{Join selectivity estimation algorithm for the \emph{less than (<)} operator re-using Algorithm \ref{alg:restriction_sel_estimation} as \(F_X\) and \(F_Y\)}
    \label{alg:join_sel_estimation}
    \KwIn{
        \\ histogram of X, \(hist_X\), is an array of length \(n_X\)
        \\ histogram of Y, \(hist_Y\), is an array of length \(n_y\)
    }
    \KwOut{Join selectivity estimation of X < Y}
    
    selectivity $\leftarrow$ 0 \;
    
    sync $\leftarrow$ merge\_sorted(hist$_X$, hist$_Y$) \; 
    
    cur\_F$_X$ $\leftarrow$ F$_X$(sync[0]) \tcp*{always zero}
    
    cur\_F$_Y$ $\leftarrow$ F$_Y$(sync[0]) \tcp*{always zero}
    
    \For{$k\gets1$ \KwTo $n_X + n_Y - 1$}{
        
        next\_F$_X$ $\leftarrow$ F$_X$(sync[k])\tcp*{using Algorithm \ref{alg:restriction_sel_estimation}}    
        
        next\_F$_Y$ $\leftarrow$ F$_Y$(sync[k])\tcp*{using Algorithm \ref{alg:restriction_sel_estimation}}
        
		selectivity += (cur\_F$_X$ + next\_F$_X$) * (next\_F$_Y$ - cur\_F$_Y$) \;
		
        cur\_F$_X$ $\leftarrow$ next\_F$_X$ \;
        
        cur\_F$_Y$ $\leftarrow$ next\_F$_Y$ \;
    }
    
    \Return{selectivity / 2}
\end{algorithm}

The goal of Algorithm \ref{alg:join_sel_estimation} is to calculate the area under the curve of the product \(F_X \cdot f_Y\), represented in Figure \ref{fig:integral_product}.
Taking the 3-bin histograms calculated in section \ref{section:hist_stats} from the running example attributes $R1.X$ and $R2.Y$, a materialized \(sync\) array would have the values [10, 15, 20, 25, 39, 45, 50] after merging, sorting and removing duplicates.
These correspond to the boundaries of the pieces in which the product \(F_X \cdot f_Y\) is linear.
Stepping through Algorithm \ref{alg:join_sel_estimation} with k from 1 to 5, corresponding to the 6 pieces represented by the sync array, we arrive at the following sum:

\begin{multline*}
P(R1.X < R2.Y) = \frac{1}{2} ( \\
    (0 + \frac{1}{6}) \times (0 - 0) +  
    (\frac{1}{6} + \frac{1}{3}) \times (\frac{1}{3} - 0) + \\ 
    (\frac{1}{3} + \frac{2}{3}) \times (\frac{8}{19} - \frac{1}{3}) +  
    (\frac{2}{3} + \frac{9}{10}) \times (\frac{2}{3} - \frac{8}{19}) + \\ 
    (\frac{9}{10} + 1) \times (\frac{28}{33} - \frac{2}{3}) +  
    (1 + 1) \times (1 - \frac{28}{33}) \\ 
) = \frac{24221}{37620} \approx 0.643833 
\end{multline*}

The correct result will have 95 rows, which corresponds to a selectivity of \(\frac{95}{144} \approx 0.659722\).
After multiplying by the cardinality of the Cartesian product of both attributes (144) and rounding the result to the nearest integer, we get the estimated row count of \(92.712 \approx 93\), which is close to the correct result size.

Figure \ref{fig:integral_product} shows a graphical depiction of the product of the CDF of $R1.X$ and the PDF of $R1.Y$, which is directly obtainable from their equi-depth histograms.
The integral in equation \ref{eq:discrete_join}, as well as the estimation calculated above using Algorithm \ref{alg:join_sel_estimation}, correspond to the area under the curve in the figure.

Figure \ref{fig:integral_product} illustrates the multiplication of the CDF(R1.X) and the PDF(R2.Y). The integral in equation \ref{eq:join}, as well as its estimation in Algorithm \ref{alg:join_sel_estimation_mcv}, correspond to calculating the area under the curve in the figure. 

Note that the code re-use in algorithm \ref{alg:join_sel_estimation} has a small performance impact. This algorithm has a time complexity of O(\( (n_X + n_Y) log(n_X + n_Y) \)) since it performs a binary search (twice) for each element of each histogram.
This binary search is not necessary since the two histograms are being scanned sequentially and the current indices are known at each iteration.
One way to avoid this overhead would be to optionally specify \(j\) as an input parameter of algorithm \ref{alg:restriction_sel_estimation}, thus reducing the time complexity to O(\(n_X + n_Y \))

\subsection{Extending to all scalar inequality operators}
Given the restriction and join selectivity estimators for the less than inequality, all scalar inequality operators can be implemented by noting the following equivalences:

Restriction selectivity:

\begin{itemize}
    \item \(P(X >= c) = 1 - P(X < c)\)
    \item \(P(X > c) = 1 - P(X < c) - P(X = c)\)
    \item \(P(X <= c) = P(X < c) + P(X = c)\)
\end{itemize}

Join selectivity:

\begin{itemize}
    \item \(P(X >= Y) = 1 - P(X < Y)\)
    \item \(P(X > Y) = P(Y < X)\)
    \item \(P(X <= Y) = P(X < Y) + P(X = Y)\)
\end{itemize}
Estimators for equality selections and joins are already implemented by almost all common systems. In case they are missing, one could assume P(X = c) and P(X = Y) to be zero, thus leading to under-/over-estimate the selectivity.

\subsection{Making Use of Other Statistics}
\label{section:other_stats}
Typically, RDBMS will collect statistics about nulls, in the form of a fraction of null values, and Most Common Values (MCV), in the form of a singleton histogram. Histograms will thus be constructed for the remaining part of the data. When the histogram statistics only refer to a fraction of the data, the methods described up to this point only provide an estimation for this fraction. The final estimation must thus take nulls and MCV into account.

As a general way to integrate such other statistics in the estimation besides the histograms, we note the following: given a non-overlapping partitioning of the data, if each partition \(j\) corresponds to a fraction \(p_j\) of the original data, and selectivity within that partition is \(s_j\), the final selectivity can be calculated by the inner product \(p \cdot s\).

Since a value is either null, a most common value, or accounted for by the histogram, the overall restriction selectivity can be calculated by the following formula:

\begin{equation}
\label{eq:overall_selectivity}
p \cdot s = p_{null} s_{null} + p_{mcv} s_{mcv} + p_{hist} s_{hist}
\end{equation}

\paragraph{Null Values} All inequality operators are strict, this means that the selectivity of null values is 0. For this reason, the first term in equation \ref{eq:overall_selectivity} is also 0.

\paragraph{Most Common Values} MCV statistics maintain pairs of values and their frequencies in the table. They are maintained for the top k frequent values, where k is a statistics collection parameter. Since MCV represent the data in its original form, it is possible to accurately compute the selectivity for these values. 

To estimate the restriction selectivity of an operator using the most common values, algorithm \ref{alg:restriction_sel_estimation_mcv} can be used. This algorithm computes the selectivity of a Boolean operator on a list of most common values by adding the frequencies of the most common values satisfying this Boolean condition.

\begin{algorithm}
    \caption{mcv\_selectivity(values, fractions, n, op), estimates the restriction selectivity, using only the MCV statistics, for a given Boolean operator op}
    \label{alg:restriction_sel_estimation_mcv}
    \KwIn{
        \\ MCV statistics of X (array of \(values\) and corresponding array of \(fractions\))
        \\ Length of MCV arrays, \(n\)
        \\ constant, \(c\)
        \\ operator, \(op\)
    }
    \KwOut{Restriction selectivity estimation of X <op> c}
    
    selectivity $\leftarrow$ 0 \;
    
    \For{$i\gets0$ \KwTo $n$}{
        \If{op(values[i], c)}{
		    selectivity += fractions[i] \;
        }
    }
    
    \Return{selectivity}
    
\end{algorithm}

For join selectivity estimation, since there is a need to combine statistics from null values, MCV, and equi-depth histograms for both X and Y, there are 9 cases that need consideration, depending on the combination of values of X and Y, as shown in Figure \ref{fig:combinedSelectivityCases}.

\begin{figure}
    \centering
    \caption{Nine cases for join selectivity estimation}
        \begin{tabular}{l|ccc}
                 & Y is null & Y in MCV & Y in histogram\\
        \hline
        X is null   & case1 & case2 & case3\\
        X in MCV    & case4 & case5 & case6\\
        X in histogram& case7 & case8 & case9\\
        \end{tabular}
    
    \label{fig:combinedSelectivityCases}
\end{figure}

For strict operators, such as inequalities, only cases 5, 6, 8, and 9 need to be calculated. This is because nulls result in empty joins. Case 9 has already been handled in Algorithm \ref{alg:join_sel_estimation}. 
For case 5, we iterate over the values in the MCV of Y. For each value, we multiply the fraction represented by that value in Y times the MCV restriction selectivity of the operator in question with the current value of the MCV of Y as the constant. This process is described in algorithm \ref{alg:join_sel_estimation_mcv}.

\begin{algorithm}[ht]
    \caption{Join selectivity estimation algorithm for any binary Boolean operator re-using algorithm \ref{alg:restriction_sel_estimation_mcv} as mcv\_selectivity(values, fractions, n, op)}
    \label{alg:join_sel_estimation_mcv}
    \KwIn{
        \\ MCV statistics of X (array of \(values_X\) and corresponding array of \(fractions_X\))
        \\ Length of MCV arrays of X, \(n_X\)
        \\ MCV statistics of Y (array of \(values_Y\) and corresponding array of \(fractions_Y\))
        \\ Length of MCV arrays of Y, \(n_Y\)
        \\ operator, \(op\)
    }
    \KwOut{Join selectivity estimation of X <op> Y}
    
    selectivity $\leftarrow$ 0\;
    
    \For{$i\gets0$ \KwTo $n$}{
        selectivity += fractions\_Y[i] * mcv\_selectivity(values\_X, fractions\_X, values\_Y[i], op) \;
    }
    
    \Return{selectivity}
    
\end{algorithm}

For cases 6 and 8, Algorithm \ref{alg:join_sel_estimation_mcv_hist} is used, by swapping the arguments. For each common value in the statistics of X, multiply its fraction by the histogram restriction selectivity of Y using the current value of X as the constant.

\begin{algorithm}[ht]
    \caption{Join selectivity estimation algorithm for the \emph{less than (<)} operator re-using algorithm \ref{alg:restriction_sel_estimation} as \(F_Y\)}
    \label{alg:join_sel_estimation_mcv_hist}
    \KwIn{
        \\ MCV statistics of X (array of \(values\) and corresponding array of \(fractions\))
        \\ Length of MCV arrays, \(n_x\)
        \\ histogram of Y, \(hist_Y\)
        \\ operator, \(op\)
    }
    \KwOut{Join selectivity estimation of X < Y}
    
    selectivity $\leftarrow$ 0 \;
    
    \For{$i\gets0$ \KwTo $n_x$}{
        selectivity += fractions[i] * F\_Y(values[i]) \;
    }
    
    \Return{selectivity}
    
\end{algorithm}

Given algorithms \ref{alg:join_sel_estimation_mcv} and \ref{alg:join_sel_estimation_mcv_hist}, the selectivity of the  \emph{less than (<)} operator considering histograms and most common values can be estimated by the following formula:

\begin{align*}
\label{eq:hist_mcv_selectivity}
hist\_mcv\_selectivity = & p_{hist_X} p_{hist_Y} s_{hist_X \times hist_Y}\\ 
+ & p_{hist_X} p_{mcv_Y} s_{hist_X \times mcv_Y} \\
+ & p_{mcv_X} p_{hist_Y} s_{mcv_X \times hist_Y} \\
+ & p_{mcv_X} p_{mcv_Y} s_{mcv_X \times mcv_Y} \\
\end{align*}

The final selectivity taking null values into account can be estimated as follows:

\begin{equation}
\label{eq:overall_join_selectivity}
selectivity = (1 - p_{null}) \times hist\_mcv\_selectivity
\end{equation}

\subsection{Implementation for Ranges and Multi-Ranges}
\label{impl:ranges}

The algorithms described above are for scalar types. An advanced type, which is implemented by many database systems is the range type. A range type is a tuple (left, right, lc, rc), where left <= right are two values of a domain with a total order. lc and rc specify whether respectively the left and right bounds are included in the range. The range type can be parameterized by the type of its bounds, e.g., range(float), range(timestamp), etc. In this section, we describe how the selectivity estimation in previous sections can be applied to the range type and the respective operators.  

PostgreSQL, as an example of DBMS that has range types, collects the statistics for range attributes in the form of two equi-depth histograms: one for the lower bounds of the ranges, and one for the upper bounds. In the following, let $X, Y$ be attributes of the same range type. Also, let $X.lower$ be the variable that represents all the lower bounds of $X$, $X.upper$ be the variable that represents all the upper bounds of $X$, and similarly for $Y$. Then it is possible to estimate the selectivity of the different range operators as follows:
\begin{itemize}
    \item $P(X << Y)= P(X.upper < Y.lower)$, where $<<$, reads strictly left of, yields true when X ends before Y starts
    \item $P(X >> Y)= P(X.lower > Y.upper)$, where $>>$, reads strictly right of, yields true when X starts after Y ends
    \item $P(X$ \&$< Y)= P(X.upper < Y.upper)$, where \&$<$, reads $X$ does not extend to the right of $Y$, yields true when $X$ ends before the end of $Y$
    \item $P(X$ \&$> Y)= P(X.lower < Y.lower)$, where \&$>$, reads $X$ does not extend to the left of $Y$, yields true when $X$ starts before $Y$ starts
    \item $P(X \&\& Y)= 1 - P(X << Y) - P(X >> Y)$, where $\&\&$ indicates the overlapping between $X$ and $Y$
    \item and so on
\end{itemize}
It is however not possible to accurately estimate the join selectivity of the operators that express total or partial containment. This is mainly because the lower and upper histograms assume Independence between the range bounds. For containment operators, we need to relate the two bounds, which explicitly breaks this assumption. 

Another consideration in estimating the selectivity of range operators is the fraction of empty ranges since these are not accounted for by the histograms. Depending on the operator, empty ranges are either always included or always excluded when compared to non-empty ranges and similarly when compared to other empty ranges.

\tododiogo{Add rules for empty ranges handling for each operator (maybe Boolean table with columns: non-empty x empty, empty x non-empty, empty x empty, and rows are operators.}

\section{Experiments}
\label{section:experiments}
This section evaluates the selectivity estimation accuracy of the proposed algorithm, and its relation to the size of the histogram statistics, i.e., the number of bins. Firstly, the proposed algorithm has been implemented in PostgreSQL 14, including support for range operators as described in section \ref{impl:ranges}. We prepared a patch, and it is currently under review for inclusion in the next release of PostgreSQL. The batch is also included as an artifact with this paper.

The experiment described in this section is thus run using our implementation in PostgreSQL 15-develop on a Debian virtual Debian machine with 32GB of Disk and 8GB of RAM. We created two relations, R1 and R2, with range attributes R1.X and R2.Y, and cardinalities of 20390 and 20060 rows respectively. Note that for range types, we use the very same algorithm as for scalar types, so the results of this experiment hold for both. 

The range values in the two relations were generated to cover a mixture of short, medium, and long ranges. We also included corner cases such as ranges with infinite bounds, empty ranges, and null values in the two relations. The following query was executed varying the number of histogram bins:
\begin{lstlisting}[language=SQL]
SELECT *
FROM R1, R2
WHERE x << y
\end{lstlisting}

Note that \verb+<<+ denotes the "strictly left of" operator, which returns true if and only if the upper bound of x is less than the lower bound of y.
PostgreSQL collects bounds histogram statistics for range attributes. 
To estimate the selectivity of the query above, histograms of the upper bound of X and lower bound of Y are used as input for Algorithm \ref{alg:join_sel_estimation}.

The optimizer statistics collector of PostgreSQL has a parameter called \emph{statistics target}, that controls the number of bins in the equi-depth histograms. The default value is 100, and it can be increased up to 10000.
The experiments described in this section were run by incrementing the statistics target by steps of 100 starting with the default value till the maximum value.

In the experiment, we observe two quantities: (1) the planning time, which is the time taken by the query optimizer to enumerate the alternative query plans, and estimate their costs, and (2) the cost estimation error defined as the absolute difference between the estimated and the actual number of rows returned
by the query, divided by the cardinality of the Cartesian product
of the two relations, which is 409013259.

\begin{figure}[ht]
\centering
\includegraphics[scale=0.45]{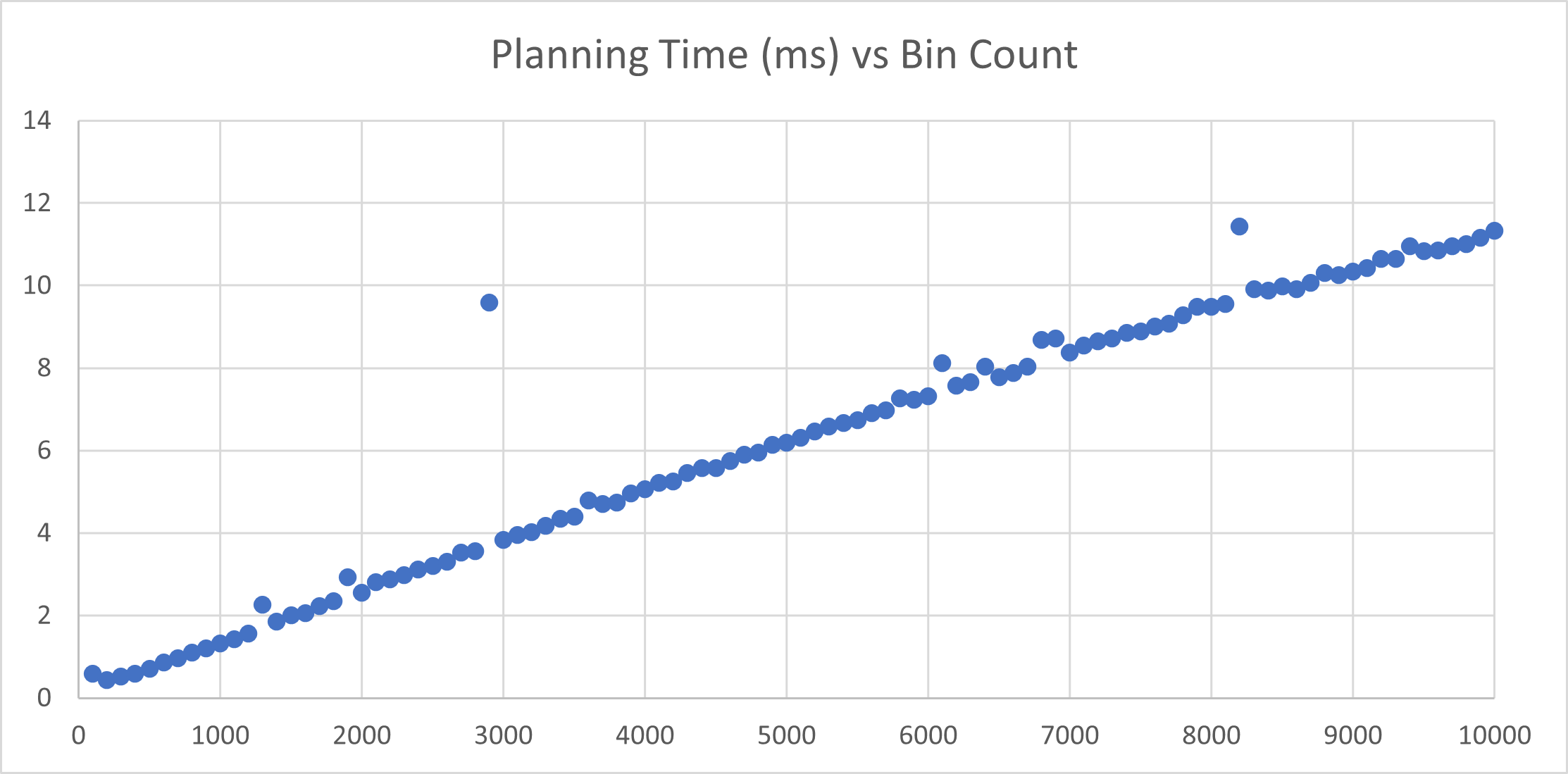}   
\caption{Query planning time (ms) V.S. the number of histogram bins in algorithm \ref{alg:join_sel_estimation}}
\label{fig:time_count}
\end{figure}

Figure \ref{fig:time_count} shows the change in planning time (in milliseconds) as the statistics target increases. Apart from two outliers, the planning time shows approximately linear behavior, indicating that the binary search does not have a significant impact on it in the allowed range of statistics targets. Recall that the analytical complexity is O(\( (n_X + n_Y) log(n_X + n_Y) \)).

\begin{figure}[ht]
\centering
\includegraphics[scale=0.45]{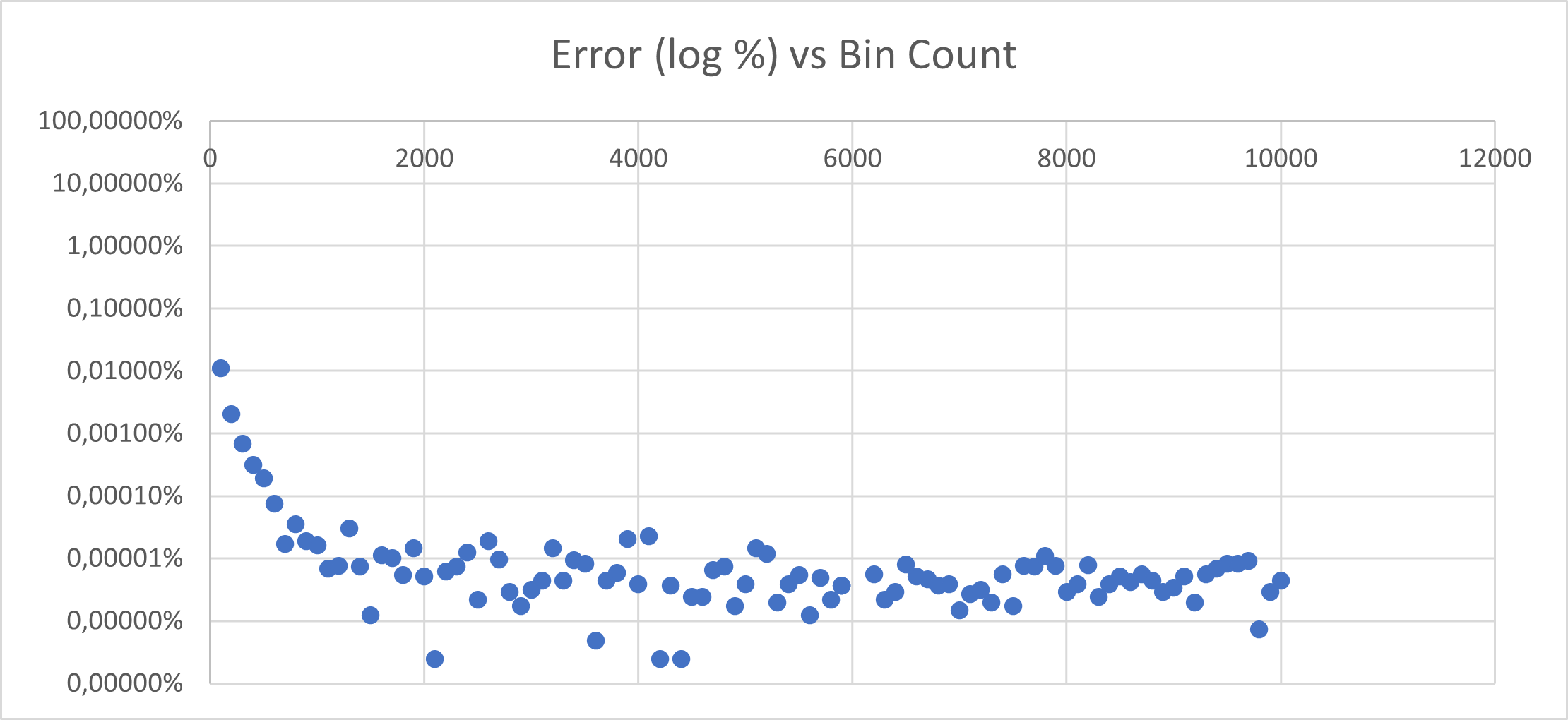}   
\caption{Log selectivity estimation error V.S. the number of histogram bins}
\label{fig:error_count}
\end{figure}

Figure \ref{fig:error_count} shows the estimation error (in a logarithmic scale) against the number of histogram bins, i.e., by varying the statistics target. As expected, using more bins leads to a lower estimation error. The largest error observed, when using only 100 histogram bins, was 1,112\%. The error then drops rapidly to less than 0,002\% at 900 bins, which corresponds to less than 5\% of each relation size. The error stays consistently around this value for the histograms bigger than 900 bins.  \diogo{Need help finding better wording for the fast decrease in error}

\begin{figure}[ht]
\centering
\includegraphics[scale=0.45]{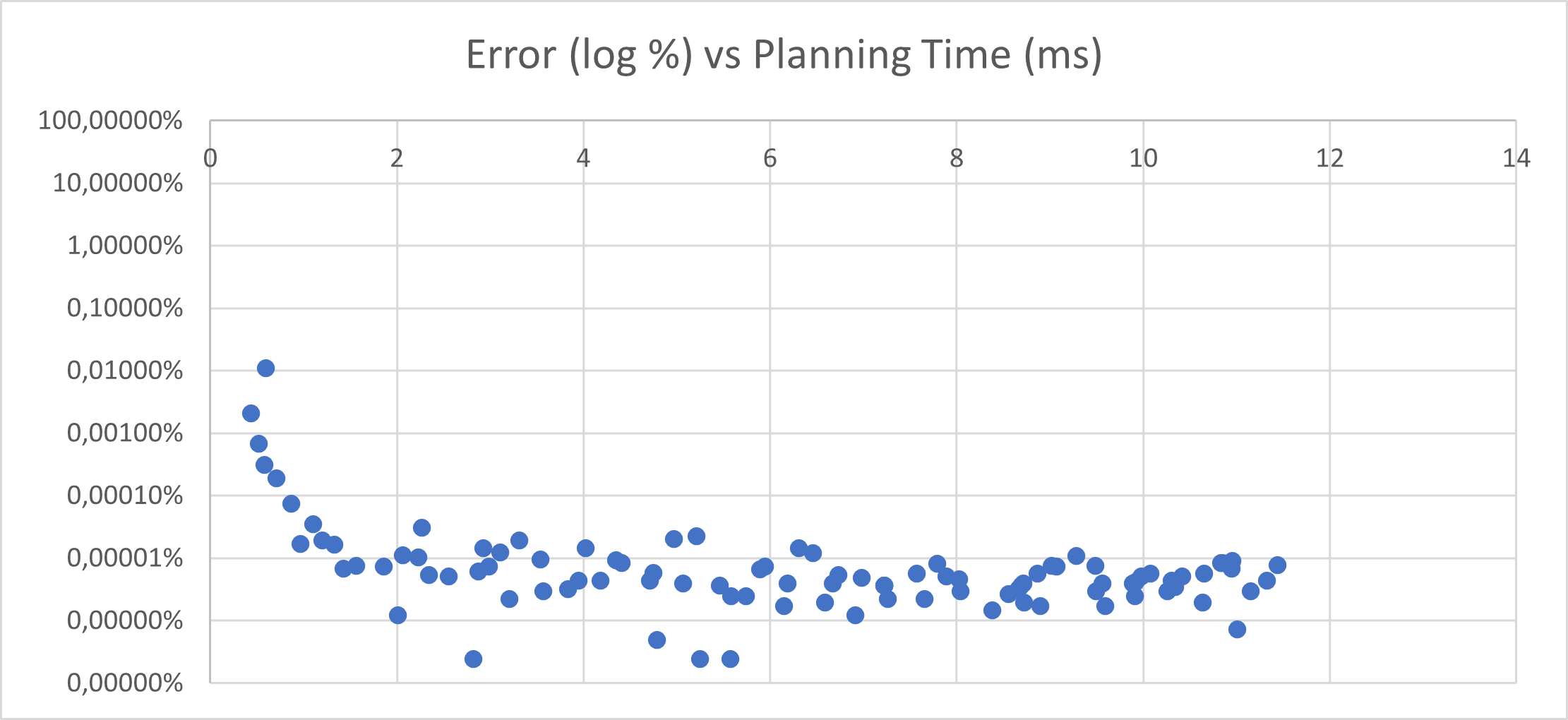}   
\caption{Log selectivity estimation error V.S. planning time (ms) }
\label{fig:error_time}
\end{figure}

Figure \ref{fig:error_time} plots the selectivity estimation error against the planning time in milliseconds. The significance of this figure is to illustrate the relation of the expenditure in terms of planning time versus the gain in terms of reduced error. This figure shows that, for the relations used, the planning time does not need to exceed 2 milliseconds to obtain extremely accurate estimations.

\section{Conclusions}
This paper proposed an algorithm for estimating the selectivity of inequality join predicates. It is a fundamental problem in databases that has not been solved before up to our knowledge. Common open-source databases, PostgreSQL and MySQL, lack implementations for this functionality. We have implemented and pushed the proposed algorithm as a patch to be included in PostgreSQL. Propitiatory databases, Oracle and SQL-Server, return fairly accurate estimations, but their algorithms are not known. Our experiments show that the proposed algorithm provides comparable estimation accuracy, slightly more accurate. To produce these estimations, the algorithm uses equi-depth histogram statistics, which are adopted in all these systems. In a practical setting, the planning time remains within 2 milliseconds, which is the accepted norm in common databases.  

\diogo{Should I write some conclusion here besides the explanation of the results above?}

\bibliographystyle{ACM-Reference-Format}
\bibliography{bib}

\end{document}